\documentstyle[preprint,eqsecnum,aps]{revtex}

\draft
\preprint{hep-lat/9312071,~~CTP\#2269}

\def\footnoterule{\kern-3pt \hrule width \hsize \kern6.2pt}

\def\pmb#1{\setbox0=\hbox{$#1$}%
\kern-.025em\copy0\kern-wd0
\kern.05em\copy0\kern-\wd0
\kern-.025em\raise.0433em\box0 }

\def\be{\begin{equation}}
\def\ee{\end{equation}}

\begin{document}

\title{Evidence for the Role of Instantons \\
in Hadron Structure from Lattice QCD\footnotemark[1]}

\footnotetext[1]{This work is supported in part by funds
provided by the U.S.~Department of Energy (D.O.E.) under contracts
\#DE-AC02-76ER03069 and \#DE-FG06-88ER40427, and the National Science
Foundation under grant \#PHY~90-13248.}

\author{M.-C.~Chu}
\address{W.~K.~Kellogg Radiation Laboratory, Caltech, 106--38\\
         Pasadena, California ~91125}
\author{J.~M.~Grandy}
\address{T--8 Group, MS B--285, Los Alamos National Laboratory\\
         Los Alamos, New Mexico ~87545}
\author{S.~Huang}
\address{Department of Physics, FM--15, University of Washington\\
         Seattle, Washington ~98195\\
         {\rm and}\\
         Center for Theoretical Physics,
         Laboratory for Nuclear Science,
         and Department of Physics\\
         Massachusetts Institute of Technology, Cambridge, Massachusetts 02139}
\author{J.~W.~Negele}
\address{Center for Theoretical Physics,
         Laboratory for Nuclear Science,
         and Department of Physics\\
         Massachusetts Institute of Technology, Cambridge, Massachusetts 02139}

\date{\today}
\maketitle

\setcounter{page}{0}
\thispagestyle{empty}

\begin{abstract}
Cooling is used as a filter on a set of gluon fields sampling the Wilson
action to selectively remove essentially all fluctuations of the gluon field
except for the instantons.  The close agreement between quenched lattice QCD
results with cooled and uncooled configurations for vacuum correlation
functions of hadronic currents and for density-density correlation functions
in hadronic bound states provides strong evidence for the dominant role of
instantons in determining light hadron structure and quark propagation in the
QCD vacuum.

\end{abstract}

\vfil
\centerline{Submitted to: {\em Physical Review D}}
\vskip1in
\hbox to \hsize{CTP\#2269 \hfil December 1993}
\eject

\section{Introduction}
\label{sec:1}

Understanding the quark and gluon substructure of hadrons is particularly
challenging because none of the standard analytic techniques of theoretical
physics is applicable to the non-perturbative solution of QCD.  Indeed, these
standard techniques don't even provide a qualitative understanding of the
mechanism responsible for the gross structure of hadrons --- whether it is the
Coulomb-like interaction between quarks arising from short wavelength
fluctuations of the gluon field, the behavior at large distances associated
with confinement, or a mechanism associated with topological structures in the
QCD vacuum corresponding in the semiclassical limit to instantons.

The goal of this work is to use lattice QCD as a tool to understand the role
of instantons in determining the gross features of the structure of hadrons
and of the propagation of quarks in the QCD vacuum.  Thus in this work, we use
lattice QCD as a means to obtain physics insight into issues which are not
experimentally accessible rather than to calculate experimental observables.

Our strategy for elucidating the role of instantons in hadron structure is to
focus on correlation functions which characterize the gross structure of
hadrons and quark propagation in the QCD vacuum and which are well described
by quenched lattice QCD calculations which sample the full Wilson action.
These calculations include all the fluctuations and topological excitations of
the gluon field and thus include the full perturbative and non-perturbative
effects of the short range Coulomb and hyperfine interactions, confinement,
and instantons.  We then use cooling as described below to remove essentially
all fluctuations of the gluon field except for the instantons which, because
of their topology, cannot be removed by local minimization of the action.
Thus, both the Coulomb interaction and confinement are almost completely
removed while retaining most of the instanton content.  To the extent to which
the gross features of hadron structure and quark propagation in the QCD vacuum
are unaffected by removing all the gluonic modes except instantons, we have
strong evidence for the dominant role of instantons.

The vacuum correlation functions we consider are the point-to-point equal-time
correlation functions of hadronic currents
$R(x)
= \langle \, \Omega \, | \, T \, J(x) \, \bar{J}(0) \, | \, \Omega \rangle$
discussed in detail in an extensive review by Shuryak \cite{ref1} and
recently calculated in quenched lattice QCD \cite{ref2}.  These correlation
functions characterize the propagation of quarks and antiquarks in all the
relevant hadronic channels and complement bound state data in the same way as
nucleon-nucleon scattering phase shifts complement deuteron properties in
characterizing the nuclear interaction.
Two results of Refs.~\cite{ref1} and~\cite{ref2} are of particular
relevance to the present work.  First, the quenched lattice calculations of
the correlation functions agree well with the dispersion relation analysis of
experimental data in all channels for which data are available (pseudoscalar,
vector, and axial vector) indicating that quenched QCD with present lattice
technology describes the gross features of quark propagation in the vacuum.
Second, the random instanton model also agrees reasonably well with the
empirical results in these channels and with the lattice results in all other
channels, indicating that if the instanton content of the QCD vacuum were
similar to that parameterized in this model, the instantons also could account
for the essential features of quark propagation.

To characterize the gross properties of hadrons, in addition to the mass, we
consider quark density-density correlation functions \cite{ref3,ref4,ref5}
$\langle h | \rho(x) \, \rho(0) | h \rangle$.
In contrast to wave functions, which have large contributions from the gluon
wave functional associated with the gauge choice or definition of a gauge
invariant amplitude \cite{ref6}, the density-density correlation function is a
gauge-invariant physical observable which directly specifies the spatial
distribution of quarks.  Comparison of these correlation functions in full
quenched QCD and retaining only the instanton content of the gluon
configuration is therefore expected to provide a quantitative indication of
the role of instantons in determining the spatial distribution of quarks in
hadrons.

This present work is strongly motivated by the physical arguments and
instanton models of QCD by Shuryak and collaborators \cite{ref7,ref8} and by
Dyakanov and Petrov \cite{ref9}.  The basic picture is that although the
dilute instanton gas approximation is inconsistent because the instanton
probability increases with its size, when the interactions of instantons are
taken into account, the vacuum may be characterized by a dense, stable
distribution of instantons.
The zero modes of massless quarks associated with
these instantons correspond to localized quark states,
and the propagation of light quarks takes place
primarily by hopping between these localized states.
In the simplest version of the model \cite{ref8}, the instanton vacuum
is characterized by a random spatial and color distribution of instantons and
anti-instantons of a single radial size $\rho$ and density $n$.  For
subsequent reference, we note that when $\rho$ is chosen to be $\frac{1}{3}$
fm to fit the vacuum gluon condensate and $n$ is chosen to be 1 fm$^{-4}$ to
fit the vacuum quark condensate, the random instanton vacuum model results in
Ref.~\cite{ref8} yield a good description of the vacuum correlation functions.

The outline of the paper is as follows.  In Section~\ref{sec:2}, we describe
the lattice calculation.  The results for the instanton content of the vacuum
are presented in Section~\ref{sec:3}, where we characterize the size and
density of instantons in cooled configurations and indicate the degree to
which other gluon fluctuations are removed.  Hadronic observables in the
cooled vacuum are presented and discussed in
Section~\ref{sec:4} and the summary and
conclusions are given in the final section.

\section{Lattice Calculations}
\label{sec:2}

As discussed in the introduction, we use cooling \cite{ref10,ref11} as a
filter to extract the instanton content of 19 gluon configurations obtained by
sampling the standard Wilson action on a $16^3 \times 24$ lattice at
$\frac{6}{g^2} = 5.7$.
A configuration is cooled by an iterated sequence of relaxation steps, where
for each step, the action is minimized locally at every link of the lattice
using the Cabibbo-Marinari \cite{ref12} algorithm
with three $SU(2)$ subgroups and $\beta = \infty$.
Although it is difficult to characterize the results of cooling precisely,
it is clear
that short wavelength, local fluctuations are removed most rapidly by cooling
steps, and topologically stabilized instanton excitations are removed most
slowly.  (Although isolated instantons can slowly shrink and eventually
disappear with extended cooling because of $O(a^2)$ errors in the lattice
action, the primary mode of removal is instanton--anti-instanton
annihilation.)
Thus, as
the number of cooling steps increases, the instanton content of the
configuration is strongly enhanced relative to all other gluon excitations.

To express this effect of cooling precisely, it is useful to write the
expectation value of an operator $O$ in uncooled and cooled configurations
as
\begin{mathletters}
\be
\langle \, O \, \rangle =
\frac{\int {\cal D}[U] \, O[U] \, e^{-S[U]}}{\int {\cal D}[U] \, e^{-S[U]}}
\label{2.1a}
\ee
and
\be
\langle \, O \, \rangle_n =
\frac{\int {\cal D} [U] \, O \left[ f_n [U] \right] \, e^{-S[U]}}
{\int {\cal D}[U] \, e^{-S[U]}}
\label{2.1b}
\ee
where $f_n[U]$ denotes the configuration of $SU(3)$ elements obtained
deterministically by applying
$n$ cooling steps to the configuration $U$.  Then, by inserting
$1=\int {\cal D}[V]\delta[V-f_n[U]]$, we may write
\begin{equation}
\langle \, O \, \rangle_n
= \frac {\int {\cal D} [V] \, O [V] \, e^{-\tilde{S}_n [V]}}
{\int {\cal D} [V] \, e^{-\tilde{S}_n [V]}}  \label{2.1c}
\end{equation}
where  $e^{-\tilde{S}_n [V]} \equiv \int {\cal D}[U]
\, \delta \left[ V - f_n [U]
\right] \, e^{-S[U]}$ is an effective action defining the distribution of
cooled configurations.  Hence, the expectation value in the cooled
configuration may be expressed
\be
\langle \, O \, \rangle_n =
\frac{\int {\cal D}[U] \, O[U] \, e^{-\left( \tilde{S}_n[U] - S[U] \right)}
\, e^{-S[U]}}{\int {\cal D}[U] \, e^{-S[U]}}
\label{2.1d}
\ee
\end{mathletters}
so that the filtering factor is seen to be
$e^{-\left( \tilde{S}_n[U]-S[U]\right)}$.  For large scale, topologically
stabilized modes present equally in $\tilde{S}$ and $S$, this filtering factor
approaches 1, while for short wavelength fluctuations effectively removed by
cooling, it approaches 0.

To monitor the filtering of different degrees of freedom as a function of
cooling steps, we measure several relevant gluonic observables.  Short
wavelength fluctuations giving rise to the
perturbative Coulomb and hyperfine interactions
are reflected in the total action $S$, and when $S$ is reduced by several
orders of magnitude, it is clear that these modes have been strongly filtered.
Denoting the action associated with a single instanton
$S_0 = \frac{8\pi^2}{g^2}$, once $\frac{S}{S_0}$ equals the number of
instantons plus anti-instantons, we know all other excitations have
essentially been removed.

Similarly, we monitor confinement by measuring the string tension extracted
from a $4\times7$ Wilson loop.
The size of the Wilson loop is relevant, since the local minimization of the
action corresponds to replacing each link by the sum of staples made up of the
other three links of each plaquette to which the original link contributes.
Thus, each cooling step replaces a Wilson loop by a bundle of loops smeared by
at most one lattice site.  So, as long as the number of cooling steps is much
smaller than the size of the loop, one must still see confinement.  However,
once the number of steps is larger than the loop size, there is nothing to
prohibit the string tension from  going to zero.  The $4\times7$ Wilson loops
were chosen as a measure of confinement for the practical reason that these
were the largest loops that were already available from an earlier calculation
at the same value of $\beta$ with the same cooling algorithm \cite{ref13}.

Finally, to monitor the instanton content, we measure the topological charge
$\langle Q \rangle$ and topological susceptibility
$\langle Q^2 \rangle$, where
\be
Q = \sum_{x_n} Q(x_n)
\ee
and we use the simplest expression for the topological charge density
\be
Q(x_n) = -\frac{1}{32\pi^2} \epsilon_{\mu\nu\rho\sigma} \, {\rm ~Re Tr~}
[U_{\mu\nu}(x_n) \, U_{\rho\sigma}(x_n) ]
\label{2.3}
\ee
Although this expression for the topological charge density is known to be
inadequate for the large fluctuations occurring  in uncooled configurations,
it is adequate for the smooth configurations which emerge
after several cooling steps.  For subsequent reference, one should note that
for a random ensemble of Poisson distributed instantons and anti-instantons,
$\langle Q \rangle = 0$ and $\langle Q^2 \rangle = I + A$,
the number of instantons plus anti-instantons.

As explained in the introduction, we characterize the propagation of quarks in
the vacuum by the vacuum correlation functions
$R(x) = \langle \Omega | T \, J(x) \, \bar{J}(0) | \Omega \rangle$.
Specifically, as in Ref. \cite{ref2},
we calculate the following correlation functions
in the indicated channels:

\noindent
\begin{minipage}[t]{3.7in}
\begin{eqnarray}
\hbox to 1truein{\rm pseudoscalar \hfil} && \nonumber\\ \noalign{\vskip-8pt}
R(x) &=& \langle \Omega | T \, J^{p}(x) \, \bar{J}^{p}(0) | \Omega \rangle
\nonumber\\
\hbox to 1truein{\rm vector \hfil} && \nonumber\\ \noalign{\vskip-8pt}
R(x) &=& \langle \Omega | T \, J_{\mu}(x) \, \bar{J}_{\mu}(0)
| \Omega \rangle \nonumber\\
\hbox to 1truein{\rm nucleon \hfil} && \nonumber\\ \noalign{\vskip-8pt}
R(x) &=& {\textstyle {1\over4}} {\rm ~Tr~} \left( \langle \Omega
| T \, J^{N}(x) \, \bar{J}^{N}(0)
| \Omega \rangle \, x_\nu \, \gamma_\nu \right) \nonumber\\
\hbox to 1truein{\rm Delta \hfil} && \nonumber\\ \noalign{\vskip-8pt}
R(x) &=& {\textstyle {1\over4}} {\rm ~Tr~}
\left( \langle \Omega | T \, J^{\Delta}_{\mu}(x) \, \bar{J}^{\Delta}_{\mu}(0)
| \Omega \rangle \, x_\nu \, \gamma_\nu \right) \nonumber
\end{eqnarray}
\end{minipage}
\hspace*{\fill}
\begin{minipage}[t]{2.7in}
\begin{mathletters}
\begin{eqnarray}
\phantom{[]} \nonumber\\ \noalign{\vskip-8pt}
     J^p &=& \bar{u} \, \gamma_5 \, d \\
\phantom{[]} \nonumber\\ \noalign{\vskip-8pt}
     J_\mu &=& \bar{u} \, \gamma_\mu \, \gamma_5 \, d \\
\phantom{[]} \nonumber\\ \noalign{\vskip-8pt}
     J^N &=& \epsilon_{abc} [u^a \, C \, \gamma_\mu \, u^b] \,
             \gamma_\mu \, \gamma_5 \, d^c \\
\phantom{[]} \nonumber\\ \noalign{\vskip-8pt}
     J_\mu^\Delta &=& \epsilon_{abc}
            [ u^a \, C \, \gamma_\mu \, u^b ] \, u^c \, .
\end{eqnarray}
\end{mathletters}
\end{minipage}

As in Refs.~\cite{ref1} and~\cite{ref2}, we consider the ratio of the
correlation function in QCD to the correlation function for non-interacting
massless quarks,
$\frac{R(x)}{R_0(x)}$,
which approaches one as $x \to 0$ and displays a  broad range of
non-perturbative effects for $x$ of the order of 1 fm.  As described in
Ref.~\cite{ref2}, the effects of lattice anisotropy are removed by calculating
$R_0(x)$ on the same lattice as $R(x)$ and measuring the ratio for a cone of
lattice sites concentrated around the diagonal.  Finite lattice volume
effects are corrected by subtracting the contributions of first images as in
Ref.~\cite{ref2}.  Finally, the correlation functions are fit
as in Ref.~\cite{ref2} by a spectral function
parameterized by a resonance mass, the
coupling to the resonance, and the continuum threshold.

Hadron density correlation functions are calculated as in Refs.~\cite{ref4}
and~\cite{ref5} for the pion, rho and
nucleon.  To avoid calculating propagators between density operators, we
consider the correlation functions
\be
\langle h | \rho_u(x) \rho_d(0) | h \rangle \propto
\int d^4y \,
\langle T \, J_h(0,T) \, \bar{u} \, \gamma_0
\, u(x+y,t) \, \bar{d} \, \gamma_0 \,
d(y,t) \, J_h(0,0) \rangle
\ee
where $J_h$ denotes a point $\pi$, $\rho$, or $N$ source
and the correlation function is averaged over the central
two time slices.  Image corrections for finite
volume effects are applied as in Ref.~\cite{ref14}.

A significant conceptual issue in comparing observables calculated
using cooled configurations with uncooled results is how to change the
renormalization of the bare mass and coupling constant as the gluon
configurations are cooled.  Clearly, as the fluctuations corresponding
to gluon exchange are filtered out, the gluonic contribution to the
physical mass and coupling constant change significantly, so our task
is to find the most physical scheme to determine the hopping parameter
$\kappa$ and the lattice spacing $a$.  A different, but equivalent
way to state the issue is to note that the cooled calculation
samples a different action, Eq.(2.1c).
For the full theory, of course,
the result should not depend on the choice of the masses or other
observables used to determine $\kappa$ and $a$.  However, after
filtering out all the gluonic excitations except instantons, there is
no reason that all physical observables should be correctly
reproduced, so different values of $\kappa$ and $a$ (as well as all
other observables) will arise from different choices of a pair of
observables to determine the parameters of the theory.  Since we are
primarily interested in the instantons after cooling, the most natural
quantity to use to determine $a$ would be the topological
susceptibility, $\chi$.  However, given the limitations of the naive
topological charge density \cite{campo},
the uncooled lattice measurement of $\chi$
is unreliable, so there would be a large uncertainty in using it to
determine $a$.  (We will see below that if we did use this
prescription, $a$ would be essentially independent of cooling.)

In the end, we have chosen to use the physical pion and nucleon masses
to determine $\kappa$ and $a$ for the cooled configurations.  As
expected, the critical kappa, $\kappa_c$ approaches the free field
value 0.125 with increasing cooling.  To the accuracy of our present
calculations there is not a significant difference between
extrapolating to $\kappa_c$ and to the physical pion mass, so results
are quite insensitive to this choice.  As will be seen below, $a$
changes by $\sim 16\%$ after 25 cooling steps when the nucleon mass is
used to set the scale, and within errors, the rho mass remains
unchanged after cooling with this value of $a$.  The other extreme
would be to keep $a$ fixed at the uncooled value and thus display
what remains in the original path integral when only instantons
are retained. This constant $a$ would also be consistent with the
constant topological susceptibility. It is a remarkable result
that these two extremes differ by only $16\%$, so that even if
one took the most conservative possible view of not changing the
scale, the qualitative results would still not be changed
significantly.

\section{Instanton content of the gluon vacuum}
\label{sec:3}

In this section we present the results of cooling the gauge field
configurations. Since we are using the quenched approximation, the
gluon configurations are not influenced by the quarks and may
therefore be fully described independently of the subsequent
discussion of hadronic observables.

To provide a clear picture of how cooling extracts the instanton
content of a thermalized gluonic configuration, we display in Fig.~1
the action density $S(1,1,z,t)$ and topological charge density
$Q(1,1,z,t)$ for a typical slice of a gluon configuration before
cooling and after 25 and 50 cooling steps. As one can see, there is no
recognizable structure before cooling.  Large, short wavelength
fluctuations of the order of the lattice spacing dominate both the
action and topological charge density.  After 25 cooling steps, three
instantons and two anti-instantons can be identified clearly.  The
action density peaks are completely correlated in position and shape
with the topological charge density peaks for instantons and with the
topological charge density valleys for anti-instantons.  Note that
both the action and topological charge densities are reduced by more
than two orders of magnitude, so that the fluctuations removed by
cooling are several orders of magnitude larger than the topological
excitations that are retained.  (Also note that because we plot
$S/\beta$ and include the factor $(32\pi^2)^{-1}$ in the definition of
$Q$ in Eq.~(\ref{2.3}), the scales in Fig.\ref{fig1} differ by
$4\pi^2/3$.)  Since instantons with sizes equal to or smaller than the
lattice spacing are strongly distorted on the lattice they can not be
distinguished from any other short wavelength fluctuations and
therefore are presumably cooled away at this point. From
Fig.\ref{fig1}(e,f) we see that further cooling to 50 steps results in
the annihilation of the nearby instanton--anti-instanton pair but
retains the well separated instantons and anti-instanton.

With this orientation from a single configuration, we now consider the
ensemble averages of observables as a function of the number of
cooling steps shown in Fig.~\ref{fig2}. As expected, the action is
dominated by the short range modes and is therefore very strongly
affected by cooling, decreasing by two orders of magnitude in the
first five steps.  The topological charge is less sensitive to short
range modes, which implies a much milder dependence on cooling.
(Note, as discussed below, that our definition of the topological
charge is only accurate after several cooling steps have
smoothed the configurations.)
At cooling step 25, the averaged total action in units of a single
instanton action is $\sim 65$ whereas $\langle Q^2 \rangle$ is $\sim
25$.  This difference indicates that there are sufficient nearby
instanton--anti-instanton pairs in each configuration that we have not
yet reached the dilute regime where $\langle Q^2 \rangle \sim A + I$.
Since the nearby pairs continue to annihilate under further cooling,
we only expect a clear plateau for the topological charge but not for
the action in this region of cooling steps. It is only when the
configurations are composed of well isolated instantons that plateaus
for both action and topological charge would start to emerge. In our
case, we expect this will happen beyond 50 cooling steps, where
$\langle S \rangle / S_0$ and $\langle Q^2 \rangle$ are nearly equal.

It should be emphasized that, because we used the naive definition for
the topological charge density operator, which suffers both additive
and multiplicative renormalizations, we cannot use $\langle
Q^2\rangle$ to estimate the topological charge susceptibility when the
configurations are not yet smooth. So the first few points in
Fig.~\ref{fig2} should not be taken too literally. On the other hand,
although the cooled configurations are indeed quite smooth and hence
do not suffer from the contamination due to short wavelength
fluctuations, the small instanton contribution is also suppressed,
given rise to potential systematic error. One may expect that
this suppression should have less effect for calculations with smaller
lattice spacing, since the running coupling constant in the instanton
action will eventually dynamically suppress small instantons.

The combined information from Figs.~\ref{fig1} and \ref{fig2} suggests
the following qualitative description of our cooled configurations.
The configurations cooled with 25 steps are comprised of smooth,
clearly recognizable instantons and anti-instantons and still retain
many nearby pairs. The configurations cooled with 50 steps consist of
more dilute instantons with their total action starting to be
dominated by the well isolated peaks.  We regard the configurations
cooled with 25 steps as providing a more complete description of the
instanton content of the original configurations, and will therefore
emphasize them in our subsequent calculation of hadronic properties.

In order to characterize the cooled configurations quantitatively
and to compare with relevant instanton models, we seek to determine
the average instanton size and the instanton--anti-instanton
density. To estimate the size we measure the topological charge density
correlation function
\begin{equation}
f(x)=\sum_y Q(y)Q(x+y)\, ,
\end{equation}
where $Q(y)$ is the topological charge density at point $y$ and the
sum is over the whole lattice. The ensemble averages of $f(x)$ at
cooling step 25 and 50 are displayed in Fig.\ref{fig3}. The strong
peak at small $x$ is the correlation of a single instanton or
anti-instanton with itself. The vanishing of $\langle f(x)\rangle$
at large $x$ implies that the topological charge
is uncorrelated at this larger distance and thus averages to zero.

If we assume that all instantons are well separated,
we would expect that each individual peak can be approximated by
the analytic instanton topological charge density
\begin{equation}
Q_\rho(x)={6\over\pi^2\rho^4}\biggl({\rho^2\over x^2+\rho^2}\biggr)^4\, ,
\label{Q0}
\end{equation}
where $\rho$ is the size parameter. To show how
$Q_\rho(x)$ depends on $\rho$ we plot $Q_\rho(x)$ in Fig.~4a
for several values of $\rho$ relevant to our present results.

Although in principle one should fit with a distribution of values of
$\rho$, a first approximation is obtained by using a single value of
$\rho$ which we will interpret as an average value.  A convolution of
$Q_\rho(x)$ with itself defines a function which can be used to fit
the lattice data with $\rho$ as the fitting parameter. The continuous
curves in Fig.~\ref{fig3} are the fitted results with $\rho=2.5a$ for
25 cooling steps and $\rho=2.8a$ for 50 cooling steps. The fits fail
to reproduce the detailed shape for $x/a \sim 5$.  Assuming a uniform
distribution for $\rho$ improved the fit somewhat, but we are
reluctant to use this method to infer the distribution.  We believe
that this apparent imperfection of the fitting is due primarily to the
nonlinear overlap of instantons as observed in Fig.\ref{fig1}.

One way to estimate the instanton density $n$, defined as the number
of instantons plus the number of anti-instantons per unit volume, is
to simply divide the total action by a single instanton action and
then divide by the space--time volume, which yields $\langle
n\rangle=64/V$ for 25 cooling steps and $\langle n\rangle=31/V$ for 50
cooling steps. In addition, once instantons in the cooled
configurations were sufficiently dilute to obey Poisson statistics we
could also estimate the density using $\langle Q^2 \rangle$, and we
note that the two estimates $\langle Q^2 \rangle$ and $\langle
S\rangle/S_0$ begin to agree after 50 cooling steps.

As a further consistency check, we have also analyzed the cooled
configurations directly by defining clusters. Two adjacent lattice
points belong to the same cluster if the product of their topological
charge density is greater than the square of a threshold value, $t$.
A threshold parameter is necessary since instantons have tails that
decay only algebraically. Once again, if instantons are dilute, the
size of the cluster for a given $\rho$ and $t$ is approximated by the
single continuum instanton formula
\begin{equation}
V_t(\rho)\equiv\int d^4x\,\,\theta[Q_\rho(x)-t]={\pi\over 2}\rho^2
\biggl[ \biggl({6\over\pi^2 t}\biggr)^{1/4}-\rho\biggr]^2 \, .
\label{Vt}
\end{equation}
We display $V_t(\rho)$ for several values of $t$ in Fig.~4b.
Note that $V_t(\rho)$ has a maximum for each threshold, and observe in Fig.~4a
how for $t=.002$ the maximum is approached as $\rho$ decreases from 3 to 2.5
to 2.
In Fig.~\ref{fig5} the distribution of $V_t$ measured
in nineteen configurations
is histogrammed for two thresholds
$t=0.003$ and $t=0.005$. One distinct feature in Fig.~\ref{fig5}
is the sharp sudden drop (indicated by arrows)
in each case beyond a maximum value $V_t^{\rm max}$, reflecting the
maxima in $V_t(\rho)$.
Since $V_t^{\rm max}$ depends on the threshold $t$,
we can use Eq.~(\ref{Vt}) to estimate its magnitude and $t$ dependence.
The ratios of these maxima at $t_1$ and $t_2$ follow the prediction of
Eq.~(\ref{Vt})
but the magnitude $V_t^{\rm max}$ is roughly
a factor of 1.5 larger in absolute value than Eq.~(\ref{Vt}).
We again attribute this discrepancy
to the overlapping between adjacent instantons.

Eq.~(\ref{Vt}) may be inverted to express $\rho$ as a
function of $V_t$,
\begin{equation}
\rho={1\over 2}\biggl[ \biggl({6\over\pi^2 t}\biggr)^{1/4}\pm
\sqrt{\biggl({6\over\pi^2 t}\biggr)^{1/2}-
4\biggl({2V_t\over\pi^2}\biggr)^{1/2}}\,\,\biggr]\, .
\label{rho}
\end{equation}
Since the cooled configurations have very few instantons with
$\rho < 2$ lattice units, we may choose the plus
sign in Eq.~(\ref{rho}) and calculate $\rho$ uniquely from $V_t$. If we
correct the overlapping problem in our lattice data by simply dividing the
lattice $V_t$ by 1.5, then we can convert Fig.~\ref{fig5} into a histogram in
terms of the sizes of instantons. Fig.~\ref{fig6} shows the result, including
the Jacobian factor $dV_t/d\rho$.  These $\rho$--histograms can be regarded as
a rough estimate of the instanton size distribution in the cooled
configurations.  At $t=0.005$, where the overlapping is less severe, we see
that the distributions are centered around the mean values determined from the
topological charge density correlation function, with widths of the order of a
half lattice spacing. Note that the factor 1.5 introduced to extract an
approximate distribution of $\rho$ does not affect our determination
of the average $\rho$ and the density $n$.

Table I summarizes our result in physical units.  The lattice constant $a$ is
determined using the proton mass which is measured as described in the next
section. The string tension in lattice units was estimated in
Reference~\cite{ref13} using Wilson loops up to sizes $7\times4$, which
corresponds to a distance of around 1 fermi.  For comparison, the relevant
parameters used in instanton models by Shuryak and collaborators \cite{ref8}
are also included.

Finally, we should note that a similar analysis of cooled configurations has
previously been carried out in the case of SU(2) with smaller lattices and
slightly different techniques \cite{ref15}.  In that work, the positions and
magnitudes of peaks in $S(x,y,z,t)$ were used to determine the distribution of
sizes of instantons.

\section{hadronic observables in the cooled vacuum}
\label{sec:4}

In this section, we present the results for quark propagation and hadron
properties in the cooled vacuum and compare them with the corresponding
results before cooling.

As in Ref.~\cite{ref2}, we extrapolate the masses and vacuum correlation
functions calculated at several values of $\kappa$ to the physical pion mass.
The masses extracted from the asymptotic decay of the correlation functions
(which agree within errors with the less accurately determined resonance
masses obtained from fitting the spectral functions) are tabulated in Table II
as a function of $\kappa$ for 25 and 50 cooling steps.  The quality of the
chiral extrapolation for masses calculated at 25 cooling sweeps is shown in
Fig.~\ref{fig7}.  We note that $M_\pi^{\,2}$, $M_\rho$, $M_N$, and $M_\Delta$
are quite linear over the relevant region of $\kappa$.  They thus provide a
good determination of the values of $a$ and $\kappa_\pi$ at which
simultaneously $M_\pi$ = 140 MeV and $M_N$ = 940 MeV.  The chiral
extrapolation at 50 steps is comparable, and together these extrapolations
yield the values for $a$ in Table I and the masses shown in Table III.  The
chiral extrapolation of the spatial dependence of the ratio of correlation
functions $\frac{R(x)}{R_0(x)}$ was carried out using polynomial extrapolation
as in Ref.~\cite{ref2}.  The quality of the extrapolation was comparable to
that shown in Fig.~\ref{fig5} of Ref.~\cite{ref2}, and is not presented here
to save space.

  At this point it is appropriate to address error estimates
for the parameters tabulated in Tables II and III. The errors
quoted in Table II for hadron masses are standard jackknife
errors. As observed in Fig.1, the magnitude of short range
fluctuations in the uncooled gluon configurations is several
orders of magnitude larger than the smooth cooled gluon fields,
which are reflected in significantly larger statistical errors
for uncooled than for cooled configurations. Hence, we
were unable to use the asymptotic decay of correlation functions
to measure hadron masses for the uncooled configurations to the
same accuracy as the cooled configurations, so in Table III the
uncooled hadron masses are determined
from the dispersion relation fit.

One source of systematic errors, as discussed in Ref.~\cite{ref2},
is our limited knowledge of the functional form of the spectral
function. Thus the resonance plus background parameterization
of the spectral function could lead to systematic errors larger
than the small statistical errors quoted in Table III.
Furthermore, since we fix the hadron mass
to its asymptotic value in the dispersion relation fit for the
cooled two-point functions, the errors for the fitted parameters
are usually underestimated, due to the nonlinear nature
of the dispersion relation fitting.

Because of these statistical and systematic errors, we emphasize
that the error bars in Table III are underestimates, and care
should be taken to avoid misinterpreting the results.
For example, one might superficially
conclude from the numbers given in Table III that the mass splitting
between the $\Delta$ and the nucleon is decreased by cooling.
However, from Fig.9 of Ref.~\cite{ref2}, we know that the
dispersion relation fitted $\Delta$ mass is about $20\%$
higher than its asymptotic mass determined by the APE group.
Therefore, in order to determine the amount of the mass
splitting between the $\Delta$ and the nucleon which originates
from perturbative one-gluon exchange, one needs to go
beyond the numerical precision of the present exploratory
calculation.

\subsection{Vacuum correlation functions of hadron currents}

The principal results for vacuum correlation functions are presented in
Figs.~\ref{fig8} and~\ref{fig9}.
As in Ref. \cite{ref2}, the individual contribution of the resonance
and continuum components of the spectral function, as well as the sum,
are plotted. In the top panels of Fig.~8, we show the
ratio of interacting to non-interacting current correlation functions,
$\frac{R(x)}{R_0(x)}$, in the pseudoscalar channel for uncooled QCD, for 25
cooling steps, and for 50 cooling steps.  This channel is by far the most
attractive of all the meson channels, as reflected in the fact that the
correlation function for interacting quarks is roughly 50 times larger than
for free quarks, and is thus the only channel to be plotted on a log scale.
Since the pion mass is used to determine the bare quark mass, masses of the
pion resonance term in Fig.~\ref{fig8} are constrained to be fixed at 140 MeV.
Note that after 25 cooling steps, the correlation function is qualitatively
similar to the uncooled result, although the magnitude at 1.5 fm is roughly
half as large.  After an additional 25 cooling steps, the peak grows in
strength.  Apparently, although the distribution of instantons after 50 steps
is more dilute and less representative of the QCD vacuum than after 25 steps,
it reproduces the uncooled correlation function slightly better.  To assure
that this behavior is not a statistical artifact, in this and every other
channel we analyzed two independent sets of 9 and 10 configurations separately
and verified that the same behavior occurred in both cases.

Analogous results for $\frac{R(x)}{R_0(x)}$ in the nucleon channel are shown
in the bottom panels of Fig.~\ref{fig8}, where again the nucleon mass is
constrained to be constant because it is used to determine the lattice
spacing.  The behavior is similar to that in the pseudoscalar channel.  After
25 sweeps, the correlation function is qualitatively similar to the uncooled
result.  In detail, the peak also appears lower after cooling, although this
time it agrees within errors.  After an additional 25 sweeps the peak height
increases again, agreeing even more closely with the uncooled result.

The ratios of correlation functions $\frac{R(x)}{R_0(x)}$ for the vector
channel are shown in the upper panels of Fig.~\ref{fig9}.  In this case, the
$\rho$ mass governing the resonance peak is unconstrained, but as seen in the
figure and in Table III, it does not change significantly with cooling.
Furthermore, in this channel there is virtually no change in the correlation
function ratio with cooling.

Finally, the ratios of correlation functions in the $\Delta$ channel are shown
in the lower panel.  Again, although the position of the $\Delta$ peak is
unconstrained, it does not shift significantly with cooling.  Although the
peak height may grow somewhat with cooling, it is also consistent within
errors with remaining constant.

The resonance masses $M$, couplings $\lambda$, and continuum thresholds $S_0$,
characterizing these correlation functions in each channel are tabulated in
Table III.  These numbers reflect the same features discussed above, and
emphasize the similarity of the results after 25 and 50 cooling steps to the
uncooled results.  In addition, one observes quite general agreement both with
phenomenological results where available and with the random instanton model
and sum rules.

\subsection{Hadron density-density correlation functions}

Density-density correlation functions in the ground state of the pion, rho,
and nucleon are shown in Fig.~\ref{fig10}.
The striking result for both the rho and the nucleon is the fact that the
spatial distribution of quarks is essentially unaffected by cooling ---
instantons alone govern the gross structure of these hadrons, as indeed they
also governed vacuum correlation functions of hadron currents in these
same channels.

The only case in which a noticeable change is brought about by cooling is in
the short distance behavior of the ground state of the pion.
This difference is understandable since in the physical pion, in addition to
instanton-induced interactions, there is also a strong attractive hyperfine
interaction arising from perturbative QCD which, combined with the 1/$r$
interaction, gives rise to the central peak in the uncooled density.  In
contrast, in the rho the hyperfine interaction has much less effect, both
because it is repulsive and because it is three times weaker.
Despite this difference at the origin, which receives small phase space
weighting, when the correlation functions are normalized to the same volume
integral as in Fig.~10, one observes that the overall size and long distance
behavior do not change appreciably with cooling.

It is noteworthy that the cooled density-density correlation functions shown
in Fig.~\ref{fig10} for the $\pi$, $\rho$, and nucleon are comparable, within
error bars.  This uniformity strongly suggests that
instantons set the overall spatial scale for these hadrons.

\section{Summary and Discussion}
\label{sec:5}

In this work, we have used cooling as an effective filter to remove most of
the excitations of the gluon field except for instantons.  For example, after
25 cooling steps, when the presence of instantons and anti-instantons is
clearly visible in the action density and topological charge density,
reduction of the action to 0.3\% of its original value has essentially
removed all the perturbative, Coulomb-like contributions and reduction of the
string tension to 27\% of its original value has removed most of the effects
of confinement.  We have shown that the instanton content of the QCD vacuum
extracted by cooling with no free parameters is remarkably similar to that of
phenomenological models for which the average instanton size
$\rho \sim \frac{1}{3}$ fm and density of the order of 1 fm$^{-4}$ are chosen
to reproduce phenomenological values of vacuum quark and gluon condensates.

We have also demonstrated nearly quantitative agreement between cooled and
uncooled vacuum
hadron current correlation functions in all channels.  Similarly we
have shown that the distribution of quarks in the ground state of the $\rho$
and nucleon, as measured by density-density correlation functions, are
virtually unchanged.  The most noticeable qualitative effect of cooling is the
removal of the peak in the pion density-density correlation at short distances
arising from the attractive hyperfine interaction, but even for the pion, the
gross size and long range behavior are not substantially altered.

The conclusion we draw  from these results is that instantons do indeed play a
dominant role in light quark propagation in the vacuum and in the low energy
structure of hadrons.  The picture which emerges is that a light quark
propagating in the QCD vacuum doesn't really respond to the details of the
huge, short-wavelength fluctuations seen in the top of Fig.~\ref{fig1}, but
rather hops between the localized quark states corresponding to the zero modes
associated with the instantons which become visible in the lower panels of
Fig.~\ref{fig1}.

Although we believe these results provide substantial evidence for the role of
instantons, we recognize several significant limitations.  Despite our best
efforts to monitor the effects of cooling, cooling remains an imprecise
filter.  In addition to exploring alternative filters and characterizing
the effect of cooling more completely it would also be worthwhile to complement
this work with a companion calculation in which one modified the Monte Carlo
algorithm to emphasize other excitations and suppress instantons.

There are several significant problems associated with use of the quenched
approximation.  Clearly, when nearly-zero modes are playing an important role
in quark propagation, it is also important to include the small weight arising
from the small eigenvalue in the determinant.  In addition, in studying the
instanton content of the vacuum, it is important to include fermion feedback
so that, for example, the tendency of quark--anti-quark pairs to bind
instanton--anti-instanton  pairs is included.  Hence, in view of the
significant role of instantons in the quenched results reported in this
work, it is important to explore the effect of dynamical fermions.

\section*{Acknowledgments}

It is a pleasure to thank Edward Shuryak for extensive discussions concerning
this work.  In addition, we acknowledge useful discussions with
Ken Johnson, Michail Polikarpov and Janos Polonyi.
We also thank the National Energy Supercomputer Center for
supercomputer resources.  This work was supported in part by funds
provided by the U.S.~Department of Energy (D.O.E.) under contracts
\#DE-AC02-76ER03069 and \#DE-FG06-88ER40427, and the National Science
Foundation under grant \#PHY~90-13248.

\begin{figure}
\caption{
Cooling history for a typical slice of a gluon configuration at fixed $x$ and
$y$ as a function of $z$ and $t$.  The left column shows the action density
$S(1,1,z,t)/\beta$ before cooling (a),
after cooling for 25 steps (c) and after 50 steps (e).
The right column shows the topological charge density $Q(1,1,z,t)$ before
cooling (b), after cooling for 25 steps (d) and after 50 steps (f).}
\label{fig1}
\end{figure}

\begin{figure}
\caption{
Mean values of three observables as a function of number of cooling steps.
(a)
Total action in units of a single instanton
action $S_0=8\pi^2/g^2$. The uncooled value
$\langle S\rangle/S_0=20,211$
is far off scale and is not plotted.
(b)
Topological charge squared.
(c)
Topological charge.}
\label{fig2}
\end{figure}

\begin{figure}
\caption{
Topological charge density--density correlation functions after 25 and
50 cooling steps.  Lattice measurements are denoted by solid points
with error bars.  The curves show the best fit obtained using a
convolution of the topological charge density for a single instanton
size $\rho$ as described in the text. The size parameters $\rho$ are
in lattice units.  }
\label{fig3}
\end{figure}

\begin{figure}
\caption{
Plots showing relevant instanton geometry.
(a) Topological charge density of a single instanton for the range of
sizes $\rho$ dominating the lattice measurements.
(b) Cluster size $V_t(\rho)$ (Eq.~\protect\ref{Vt})
for the single instanton as a function of its size $\rho$ at
several values of threshold $t$.
}
\label{fig4}
\end{figure}

\begin{figure}
\caption{
Distribution of instantons as a function
of cluster size. The histograms show the observed counts
for given $V_t$ with a bin size 5
using 19 configurations.
The number of cooling steps and value of the threshold $t$ are shown in
each plot. The arrows indicate the sudden drop at the value
$V_t^{\rm max}$ in each case,
as discussed in the text.
}
\label{fig5}
\end{figure}

\begin{figure}
\caption{
Distribution of instantons as a function of $\rho$.
Each bin in the histogram of Fig.~5
is converted from $V_t$ to $\rho$ using Eq.~\protect\ref{rho} and
multiplied by the Jacobian $dV_t/d\rho$.}
\label{fig6}
\end{figure}

\begin{figure}
\caption{
Chiral extrapolation of hadron masses for configurations with 25 cooling
steps.  Masses in lattice units calculated at four values of $\kappa$ are
denoted by error bars.  The linear extrapolations of $M_\pi^{\,2}$ to
determine $\kappa_c = 0.1285$ and of $M_\rho$, $M_N$, and $M_\Delta$ to the
point at which $M_\pi$ = 140 MeV are shown by the straight lines.  }
\label{fig7}
\end{figure}

\begin{figure}
\caption{
Comparison of uncooled and cooled vacuum correlation function ratios,
$\frac{R(x)}{R_0(x)}$, for pseudoscalar currents $(P)$ and nucleon currents
$(N)$.  The left, center, and right panels show results for uncooled QCD, 25
cooling steps, and 50 cooling steps respectively.  The solid points with error
bars denote lattice correlation functions extrapolated to $M_\pi$ = 140 MeV.
The solid lines denote fits to the correlation functions using a
three-parameter spectral function, and the dashed and dotted curves show the
contributions of the continuum and resonance components of the spectral
functions respectively.  The upper scale shows the spatial separation in
lattice units and the lower scale shows the separation in physical units using
the values of $a$ in Table I determined from the nucleon mass.  }
\label{fig8}
\end{figure}

\begin{figure}
\caption{
Comparison of uncooled and cooled vacuum correlation function ratios,
$\frac{R(x)}{R_0(x)}$, for vector currents $(V)$ and Delta currents $(D)$.
The notation is the same as in Fig.~8.  }
\label{fig9}
\end{figure}

\begin{figure}
\caption{
Comparison of uncooled and cooled density-density correlation
functions for the pion, rho, and nucleon.  The solid circles denote
the correlation functions calculated with uncooled QCD, the open
circles with error bars show the results for 25 cooling steps, and the
crosses denote the results for 50 cooling steps.  The rho and pion
results are compared for $M_\pi^{\,2}=0.16$ GeV$^2$ and the nucleon
results are compared for $M_\pi^{\,2}=.36$ GeV$^2$.  As in
Figs.~\protect\ref{fig8} and~\protect\ref{fig9}, the separation is
shown in physical units using values of $a$ from Table~I.  All
correlation functions are normalized to 1 at the origin, except for
the cooled pion correlation functions, which are normalized to have
the same volume integral as the uncooled pion result.  Errors for the
uncooled results and for 50 steps, which have been suppressed for
clarity, are comparable to those shown for 25 steps.}
\label{fig10}
\end{figure}

\begin{table}
\caption{Summary of properties of cooled configurations}
\medskip
\begin{tabular}{ccccccc}
Cooling steps & $\langle S\rangle/S_0$ & $\sigma a^2$ & $a$ (fm)
& $\rho$ (fm) & $n$ (${\rm fm}^{-4}$) & $\chi$ (${\rm MeV}^4$) \\
\tableline
0 & 20,211 & 0.18 & 0.168 &      &      &           \\
25& 64     & 0.05 & 0.142 & 0.36 & 1.64 & $(177)^4$ \\
50& 31     & 0.03 & 0.124 & 0.35 & 1.33 & $(200)^4$ \\
\tableline
Instanton Model & & & & 0.33 & 1.0 & $(180)^4$ \\
\end{tabular}
\label{table1}
\end{table}

\begin{table}
\vspace*{0.4in}
\caption{Hadron masses in lattice units as a function of $\kappa$ for 25 and
50 cooling steps.  The extrapolated values of $\kappa_c$ are 0.1285(5) for 25
steps and 0.1283(5) for 50 steps.}
\medskip
\begin{tabular}{lcccc}
\multicolumn{5}{l}
{\bf 25 Cooling Steps {\vrule width0pt depth8pt}} \\
\tableline
{\vrule width0pt depth10pt}
$\kappa$ & $M_\pi a$ & $M_\rho a$ & $M_N a$ & $M_\Delta a$ \\
0.122 \qquad\qquad & 0.630(13) & 0.713(13) & 1.089(21) & 1.102(17) \\
0.124        & 0.513(16) & 0.631(16) & 0.958(23) & 0.987(19) \\
0.1255       & 0.416(20) & 0.572(17) & 0.862(29) & 0.911(22) \\
0.127        & 0.301(28) & 0.507(16) & 0.750(44) & 0.839(33) \\
\noalign{\vspace*{0.4in}}
\tableline
\tableline
\multicolumn{5}{l}
{\bf 50 Cooling Steps {\vrule width 0pt depth9pt}} \\
\tableline
{\vrule width0pt depth10pt}
$\kappa$     & $M_\pi a$ & $M_\rho a$ & $M_N a$ & $M_\Delta a$ \\
0.122 \qquad\qquad & 0.571(12) & 0.662(16) & 0.999(17) & 1.005(29) \\
0.124        & 0.454(13) & 0.581(15) & 0.874(22) & 0.896(26) \\
0.127        & 0.268(39) & 0.471(28) & 0.603(63) & 0.715(50) \\
\end{tabular}
\label{table2}
\end{table}

\begin{table}
\renewcommand{\baselinestretch}{1.4}
\caption{Hadron Parameters determined from point-to-point vacuum
correlation functions for uncooled and cooled configurations.}
\begin{tabular}{lllcc}
Channel & Source & M (GeV) & $\lambda$ & $\sqrt{s_0}$ (GeV)\\
\tableline
Vector & lattice (cool=00)
       & $0.72\pm 0.06$ & $(0.41\pm 0.02\,GeV)^2$ & $1.62\pm 0.23$\\
       & lattice (cool=25)
       & $0.65\pm 0.03$ & $(0.385\pm 0.004\,GeV)^2$ & $1.38\pm 0.05$\\
       & lattice (cool=50)
       & $0.70\pm 0.05$ & $(0.410\pm 0.005\,GeV)^2$ & $1.42\pm 0.04$\\
       & instanton\tablenotemark[1] &
         $0.95\pm 0.10$ & $(0.39\pm 0.02\, GeV)^2$ & $1.50\pm 0.10$ \\
       & phenomenology\tablenotemark[2] &
         0.78 & $(0.409\pm 0.005GeV)^2$ & $1.59\pm 0.02$ \\
\tableline
Pseudoscalar & lattice (cool=00)
       & $0.156\pm 0.01$ & $(0.44\pm 0.01\,GeV)^2$ & $<1.0$\\
       & lattice (cool=25) & $0.140\tablenotemark[4]$
       & $(0.341\pm 0.010\,GeV)^2$ & $1.05\pm 0.15$\\
       & lattice (cool=50) & $0.140\tablenotemark[4]$
       & $(0.475\pm 0.015\,GeV)^2$ & $1.80\pm 0.18$\\
       & instanton\tablenotemark[1] &
         $0.142\pm 0.014$ & $(0.51\pm 0.02\,GeV)^2$ & $1.36\pm 0.10$ \\
       & phenomenology\tablenotemark[2] &
         0.138 & $(0.480 GeV)^2$ & $1.30\pm 0.10$ \\
\tableline
Nucleon & lattice (cool=00)
       & $0.95\pm 0.05$ & $(0.293\pm 0.015\,GeV)^3$ & $<1.4$\\
       & lattice (cool=25) & $0.938\tablenotemark[5]$
       & $(0.281\pm 0.004\,GeV)^3$ & $1.47\pm 0.13$\\
       & lattice (cool=50) & $0.938\tablenotemark[5]$
       & $(0.297\pm 0.004\,GeV)^3$ & $1.54\pm 0.11$\\
       & instanton\tablenotemark[1] &
         $0.960\pm 0.030$ & $(0.317\pm 0.004\,GeV)^3$ & $1.92\pm 0.05$ \\
       & Sum Rule\tablenotemark[3] &
         $1.02\pm 0.12$ & $(0.324\pm 0.016\,GeV)^3$ & $1.5$ \\
       & phenomenology\tablenotemark[2] &
         0.939 & ------ & $1.44\pm 0.04$ \\
\tableline
Delta & lattice (cool=00)
       & $1.43\pm 0.08$ & $(0.326\pm 0.020\,GeV)^3$ & $3.21\pm 0.34$\\
       & lattice (cool=25)
       & $1.06\pm 0.04$ & $(0.285\pm 0.002\,GeV)^3$ & $1.91\pm 0.08$\\
       & lattice (cool=50)
       & $1.05\pm 0.09$ & $(0.298\pm 0.003\,GeV)^3$ & $2.22\pm 0.06$\\
       & instanton\tablenotemark[1] &
         $1.440\pm 0.070$ & $(0.321\pm 0.016\,GeV)^3$ & $1.96\pm 0.10$ \\
       & Sum Rule\tablenotemark[3] &
         $1.37\pm 0.12$ & $(0.337\pm 0.014\,GeV)^3$ & $2.1$ \\
       & phenomenology\tablenotemark[2] &
         1.232 & ------  & $1.96\pm 0.10$ \\
\end{tabular}
\tablenotetext[1]{Instanton Liquid Model by Shuryak et al.}
\tablenotetext[2]{Phenomenology estimated by Shuryak
and from the particle data book.}
\tablenotetext[3]{QCD sum rule by Belyaev and Ioffe [17].}
\tablenotetext[4]{Used to fix the quark mass.}
\tablenotetext[5]{Used to fix the lattice constant.}
\label{table3}
\end{table}

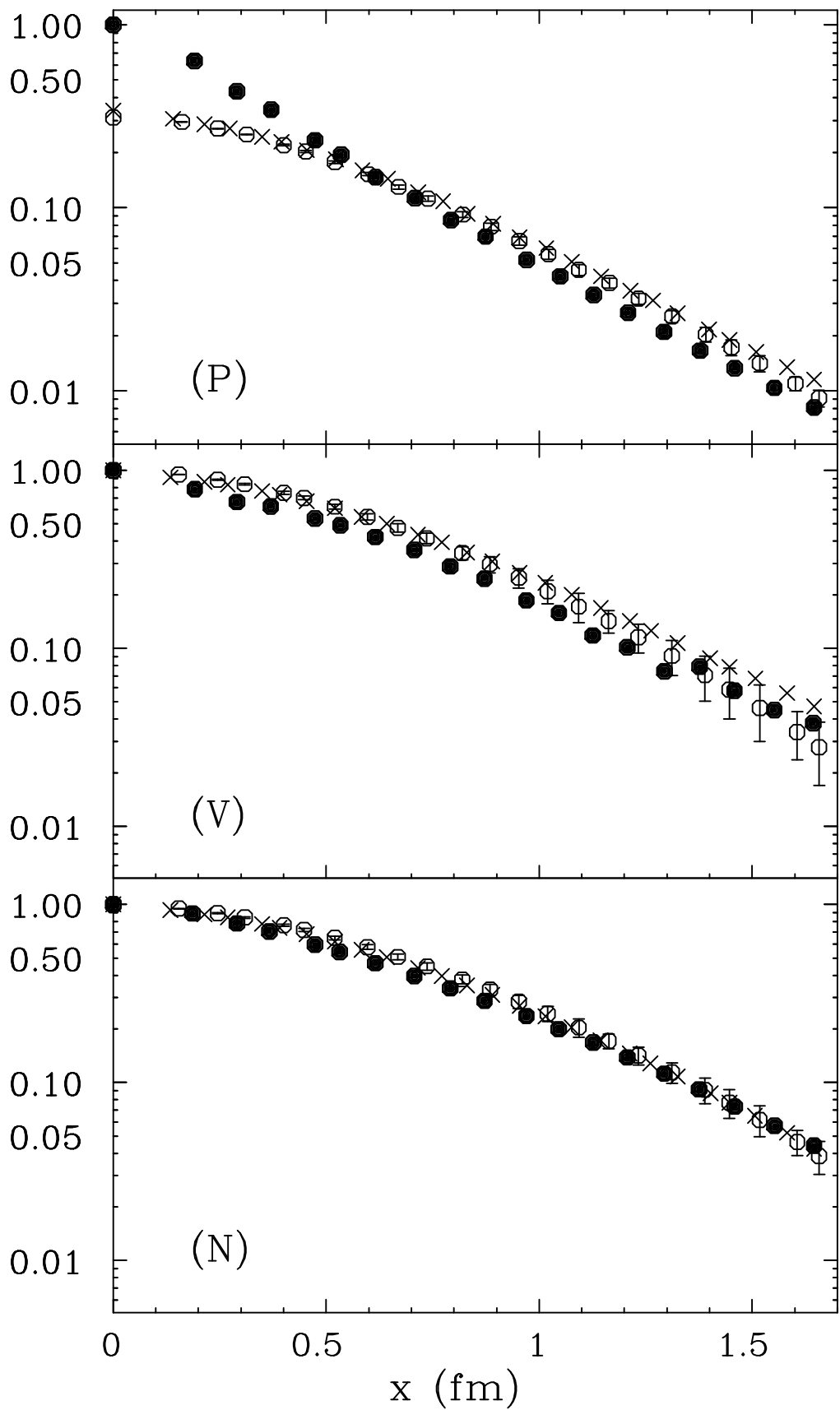
\end{document}